\documentstyle[floats,eqsecnum,preprint,aps,psfig]{revtex}
\tightenlines
\def\Jp{{J'}}
\def\Jd{{J_{\times}}}
\def\Sv{\vec{S}}
\def\Tv{\vec{T}}
\def\Sz{S^z}
\def\Tz{T^z}
\def\Mag{\langle M \rangle}
\def\hc#1{h_{c_{#1}}}
\def\ie{{\it i.e.}}
\def\beq{\begin{equation}}
\newcommand{\eeq}[1]{\label{#1} \end{equation}}
\def\bea{\begin{eqnarray}}
\def\eea{\end{eqnarray}}
\def\nn{\nonumber}
\font\aefont=msam10

\def\ale{\,\hbox{\aefont\char"2E}\,}
\draft
\begin{document}
\preprint{
\begin{minipage}[t]{\columnwidth}
\rightline{to appear in Eur.\ Phys.\ J.\ {\bf B} \hfill cond-mat/9910438}
\rightline{ETH-TH/99-28}
\rightline{}
\end{minipage}
}
\title{Magnetization plateaux and jumps in a class of 
frustrated ladders: A simple route to a complex behaviour}
\author{A.\ Honecker$^{(a)}$\footnote{
A Feodor-Lynen fellow of the Alexander von Humboldt-foundation. \\
Present address: Institut f\"ur Theoretische Physik, TU Braunschweig,
     38106 Braunschweig, Germany.},
F.\ Mila$^{(b)}$, M.\ Troyer$^{(a)}$}
\address{
$^{(a)}$ Institut f\"ur Theoretische Physik, ETH-H\"onggerberg,
     8093 Z\"urich, Switzerland \\
$^{(b)}$ Laboratoire de Physique Quantique, Universit\'e Paul Sabatier,
     31062 Toulouse Cedex, France \\
}
\date{October 27, 1999; revised December 21, 1999}
\maketitle

\begin{abstract}
We study the occurrence of plateaux and jumps in the magnetization curves of a
class of frustrated ladders for which the Hamiltonian can be written in terms of
the total spin of a rung. We argue on the basis of exact diagonalization of
finite clusters that the ground state energy as a function of 
magnetization can be obtained as the minimum - with Maxwell constructions if 
necessary - of the energies of a small set of spin 
chains with mixed spins. This allows us to predict with very elementary methods 
the existence of plateaux and jumps in the
magnetization curves in a large parameter range, and to provide very 
accurate estimates of these magnetization curves from exact or DMRG results for
the relevant spin chains.
\end{abstract}

\vspace{0.2cm}
\pacs{PACS numbers: 75.10.Jm, 75.40.Cx, 75.45.+j, 75.60.Ej}

\section{Introduction}

It is by now well established that the magnetization curve of a 
low-dimensional magnet does not always correspond to 
a smooth increase between zero magnetization and 
saturation but can exhibit plateaux at some rational values of the
magnetization (see e.g.\ \cite{NiMi,AOY,Totsuka,CHPPRL,CHPPRB,Mila,CaGy,MiUe}).
The experimental investigation of this effect has attracted a lot of
attention recently (see e.g.\ \cite{CHABetal,NHSKNT,STKTUOTMG,srcu2bo3}), and
frustrated systems emerge as prominent candidates. 
In fact, a number of
papers have convincingly demonstrated that frustrated systems can indeed
exhibit plateaux. However, in the analysis of any specific model, the
proof of the existence of a plateau usually relies on quantum field theory
methods, while the actual calculation of the magnetization curves is
performed e.g.\ via exact diagonalizations of finite clusters. In that respect,
models that allow for a simpler and unified analysis would be welcome.

In this paper, we analyze a class of models for which precise calculations can
be performed, and for which the occurrence of plateaux and jumps can be explained
in very simple terms and, to a certain extent, proved. These models are a class
of $N$-leg $S=1/2$ ladders in an external magnetic field $h$ described by the 
Hamiltonians:
\beq
H = J \sum_{x=1}^L \left(\sum_{i=1}^N \Sv_{i,x}\right)
             \cdot \left(\sum_{j=1}^N \Sv_{j,x+1}\right)
   + \Jp \sum_{x=1}^L {1 \over 2} \left(\left(\sum_{i=1}^N \Sv_{i,x}
        \right)^2-{3 N \over 4}\right)
   - h \sum_{x=1}^L \sum_{i=1}^N \Sz_{i,x} \, .
\eeq{Hop}
The particularity of these ladders is that the Hamiltonian depends only on the
total spin of each rung $\Tv_x=\sum_{i=1}^N \Sv_{i,x}$.
As a consequence, the total spin of a rung is a good quantum number, and the
eigenvalues of $H$ can be classified according to the value $T_x$ of the total 
spin of each rung. In other words, the diagonalization of $H$ is equivalent to
diagonalizing the family of Hamiltonians $H(\{T_x\})$
 \beq
H(\{T_x\}) = J \sum_{x=1}^L \Tv_x 
             \cdot \Tv_{x+1}
   + \Jp \sum_{x=1}^L {1 \over 2} \left( \Tv_x
       ^2-{3 N \over 4}\right)
   - h \sum_{x=1}^L \Tz_x \, .
\eeq{Heff}
where $\Tv_x^2=T_x(T_x+1)$, and $T_x=N/2,N/2-1,...$. So the problem is
equivalent to spin chains in a magnetic field with different values of 
the spin at each site. 

\begin{figure}[t]
\centerline{\psfig{figure=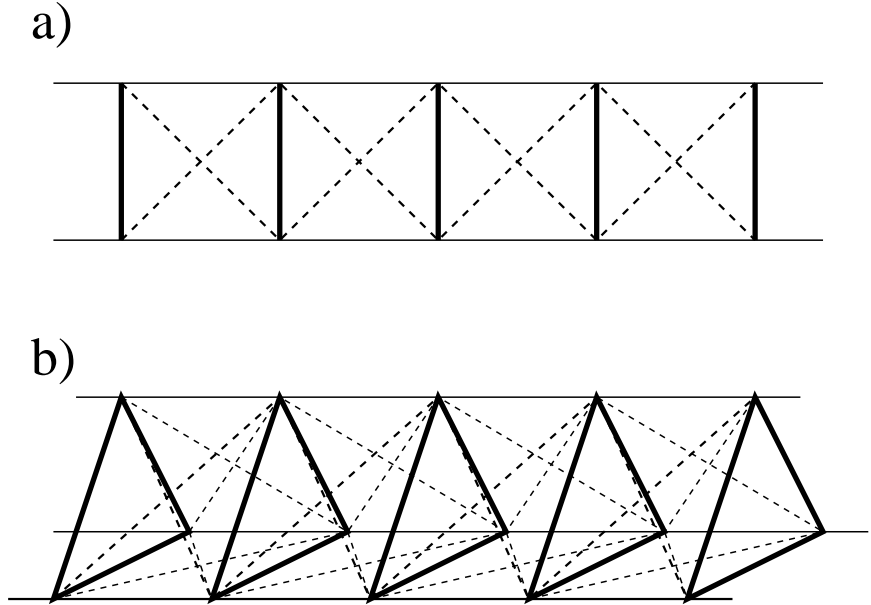,width=0.7\columnwidth}}
\bigskip
\caption{
Geometry of the a) two-leg and b) three-leg ladders considered in the
present paper. The thick full lines denote coupling $\Jp$, the
thin full lines coupling $J$ and the dashed lines coupling $\Jd$.
Throughout this paper we use $J = \Jd$. 
\label{ladders}
}
\end{figure}

Although the main ideas of the analysis could be extended to the general case,
we have decided for the sake of simplicity to consider only the cases $N=2$ and
$3$ in this paper.
For $N=2$, this model can be considered as the $J=\Jd$ special case of a
two-leg ladder with an additional diagonal coupling $\Jd$
(see Fig.\ \ref{ladders}a), \ie\
\bea
H &=& J \sum_{x=1}^L \left(\Sv_{1,x} \cdot \Sv_{1,x+1}
                       + \Sv_{2,x} \cdot \Sv_{2,x+1} \right)
   + \Jp \sum_{x=1}^L \Sv_{1,x} \cdot \Sv_{2+1,x} \nn \\
 &&+ \Jd \sum_{x=1}^L \left(\Sv_{1,x} \cdot \Sv_{2,x+1}
                       + \Sv_{2,x} \cdot \Sv_{1,x+1} \right)
   - h \sum_{x=1}^L \sum_{i=1}^2 \Sz_{i,x} \, .
\label{HopDiag2}
\eea
The ground state phase diagram of the Hamiltonian (\ref{HopDiag2})
has been studied extensively using DMRG \cite{LFS,LeSo,Wang}, series
expansions \cite{WKO}, matrix product states \cite{KoMi} and bosonization
\cite{KiSo,AEN}.

Similarly for $N=3$, the model (\ref{Hop}) arises as the $J=\Jd$ special case
of the cylindrical three-leg ladder with additional diagonal couplings
shown in Fig.\ \ref{ladders}b) whose Hamiltonian is given by
\bea
H &=& J \sum_{x=1}^L \sum_{i=1}^3 \Sv_{i,x} \cdot \Sv_{i,x+1}
   + \Jp \sum_{x=1}^L \sum_{i=1}^3 \Sv_{i,x} \cdot \Sv_{i+1,x} \nn \\
 &&+ \Jd \sum_{x=1}^L \sum_{i=1}^3 \Sv_{i,x} \cdot
        \left(\Sv_{i-1,x+1} + \Sv_{i+1,x+1}\right)
   - h \sum_{x=1}^L \sum_{i=1}^3 \Sz_{i,x} \, .
\label{HopDiag3}
\eea
Ground state properties of similar three- (and four-) leg ladders have
been investigated recently in \cite{GhoBo,ALN}.

To understand why it is much simpler to study these specific frustrated
ladders, we first note that one way to calculate the magnetization as a
function of the field consists in first calculating the ground state
energy as a function of magnetization $\Mag=2 \Tz_{{\rm tot}} / (LN)$
(where $\Tz_{{\rm tot}} = \sum_{x=1}^L \Tz_x$) for zero
field. The magnetization curves can then be constructed from this using the
identity $h = {\partial E}/{\partial \Tz_{{\rm tot}}}$
\footnote{Since the magnetic field is coupled to a conserved quantity in
(\ref{Hop}) and (\ref{Heff}), one has $E(\Tz_{{\rm tot}},h) =
E(\Tz_{{\rm tot}},0) - h \Tz_{{\rm tot}}$. From this one obtains
a finite-size formula for the transition between ground states in
$\Tz_{{\rm tot}}$ and $\Tz_{{\rm tot}}+1$:
$h  = E(\Tz_{{\rm tot}}+1,0) - E(\Tz_{{\rm tot}},0)$.
}.
It turns out that, for the present model, the ground state energy as a 
function of magnetization does {\it not} require a calculation of the ground
state energy in all sectors $\{T_x\}$ but can be deduced from a few simple 
sectors only. Of course, this is useful
numerically because these sectors are simpler to study than the original model,
but the main advantage is qualitative since it leads to a simple interpretation
of the accidents - plateaux and jumps - of the magnetization curve, which are 
all related to level crossings between different sectors. 

\section{The two-leg case}

First we consider the case of two legs. This case is particularly simple because
a large number of eigenstates can be constructed exactly. The basic idea is the
following: The total spin of each rung can be 0 or 1 in that case. If it is
0, then there is no coupling with the neighbouring rungs (for the present
model this appears to have been noticed first in \cite{BoGa}). So any state
with $N_t$ spatially separated triplets on the rungs in a sea of singlets is an
eigenstate of the Hamiltonian (\ref{Hop}) with energy
\beq
E_{st}(N_t) = -{3 \over 4} \Jp L + \Jp N_t \, ,
\eeq{Eld}
($N_t \le L/2$ due to the condition of spatial separation).
The lowest energy among (\ref{Eld}) for a given magnetization
$\Mag$ is found when all triplets are polarized, \ie\ for the smallest
possible $N_t$. Then one has $\Mag = N_t/L$. With fully polarized
triplets one can also construct exact eigenstates of the Hamiltonian
(\ref{Hop}) for $L/2 \le N_t \le L$. Namely one puts one fully
polarized triplet every second rung and fills the remaining polarized
triplets in the remaining rungs. The energy of such a state is
$E_{st}(N_t) = -{3 \over 4} \Jp L + \Jp N_t + (2 N_t -L) J$.
To summarize, we give the formula for the lowest energy
among these exact eigenstates for a given magnetization $\Mag$
\beq
E_{st}(\Mag) = \cases{
\displaystyle
\left(-{3 \over 4} + \Mag\right) \Jp L
    & for $\displaystyle 0 \le \Mag \le {1 \over 2}$, \cr
\displaystyle
\left(-{3 \over 4} + \Mag\right) \Jp L + 2 \left(\Mag -{1 \over 2}\right) JL
    & for $\displaystyle {1 \over 2} \le \Mag \le 1$.}
\eeq{Eexact}

In the limit where $\Jp\gg J$ the ground state is expected to be found among
these states for any magnetization since this is the only way to minimize the
number of triplets. However if $\Jp$ is small enough - and certainly if it were
negative and large - it will be more favourable to put triplets everywhere
because the energy gain due to fluctuations between neighbouring triplets will
dominate. The energy obtained by 
putting triplets (not necessarily polarized) on
{\it each} rung is given by
\beq
E_{tt} = {1 \over 4} \Jp L + J E^{S=1}(L,S^z) \, ,
\eeq{Ett}
where $E^{S=1}(L,S^z)$ is the energy of an $S=1$ chain with coupling
constant $1$, length $L$ and a given $S^z$-sector.

We have checked for finite ladders using exact diagonalizations
for up to 24 sites in total ($L=12$ in (\ref{HopDiag2}))
that the states corresponding to
(\ref{Eexact}) and (\ref{Ett}) are in fact the only ground states which
arise in an external magnetic field $h$ for antiferromagnetic $J,\Jp > 0$, 
apart from special values of the magnetic field 
for which there seems to be a jump in the magnetization.
Then assuming that this remains true in the thermodynamic limit,
a very simple discussion of the magnetization curve can be given.
The starting point is the energy as a function of magnetization for the
relevant states, namely $E_{st}$ and $E_{tt}$. For $E_{st}$
(Eq.\ (\ref{Eexact})) we have analytic expressions. For $E_{tt}$
(Eq.\ (\ref{Ett})), we need the magnetization curve of a spin-1 chain. Since
values were only quoted in the literature for rather small systems
\cite{PaBo,SaTa}, we have computed it
for periodic boundary conditions and $L \le 60$ using
White's DMRG method \cite{White,DMRGbook}. In all computations we have
performed around 30 sweeps at the target system size increasing the
number of kept states up to $m=400$ during the final sweeps. The
large number of sweeps was necessary because of the choice of
periodic boundary conditions for the chains. 
To test the reliability of our calculation, we have compared our estimates
of the ground state energy per site $e_{\infty}$ and of the  gap to 
magnetic excitations $\Delta_{\infty}$ with available results.
An estimate for these two quantities is obtained by applying
a Shanks transformation to our data for $L=20$, $40$ and $60$:
$e_{\infty} = -1.401484(5)$ and $\Delta_{\infty} = 0.4106(2)$.
Although we did not try to push the calculation as far as possible since it had
to be repeated for all magnetizations, 
these estimates compare well with the estimates 
obtained in earlier works which were aiming at as high accuracy
as possible: Ref.\ \cite{WhHu} obtained
$e_{\infty} = -1.401484038971(4)$, $\Delta_{\infty} = 0.41050(2)$
using DMRG and ref.\ \cite{GJL} estimated 
$e_{\infty} = -1.401485(2)$, $\Delta_{\infty} = 0.41049(2)$
by exact diagonalization of chains with length up to $L=22$ sites.
In the following, we will use the results for 60 sites without any
extrapolation. This gives an approximation of the
ground state energy of an infinite chain with an accuracy of $10^{-4}$, 
which is more than enough for the present discussion.

\begin{figure}[t]
\psfig{figure=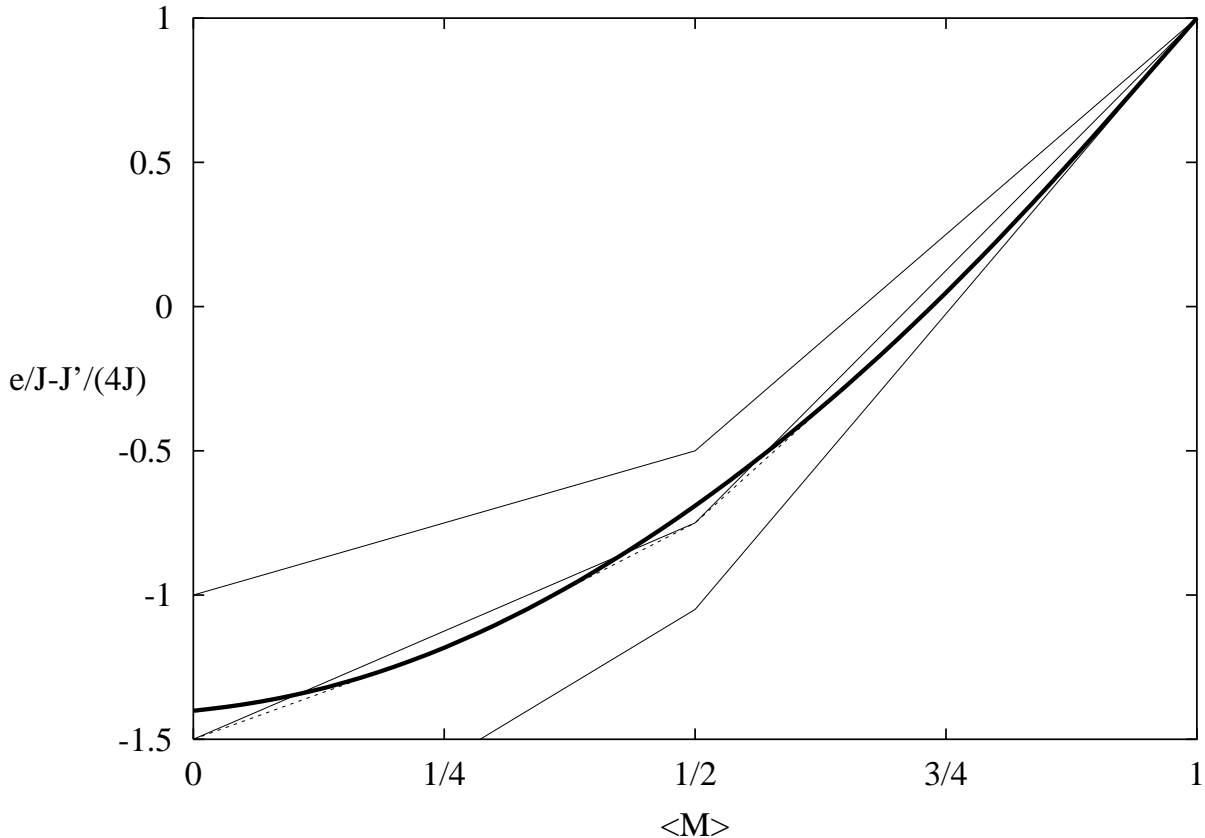,width=\columnwidth,angle=270}
\smallskip
\caption{
Ground state energy per site versus magnetization.
Bold line: Spin-1 chain Eq.\ (\ref{Ett}). Thin lines: Eq.\
(\ref{Eexact}) for $\Jp/J = 1$, $1.5$ and $2.1$ (from top
to bottom). Dashed lines: Maxwell constructions required for
$\Jp/J = 1.5$.
\label{energy2}
}
\end{figure}

The results for the energy as a function of magnetization
are plotted in Fig.\ \ref{energy2} for different values of $\Jp/J$. 
To make the
comparison between the different cases easier, we have shifted the energies by
$\Jp/(4J)$ so that $E_{tt}$ is independent of $\Jp$. 
Then the result depends only on the position of $E_{st}$ with
respect to $E_{tt}$.

If $\Jp > 2 J$, $E_{st}$ is always below $E_{tt}$.
The energy is then a piece--wise linear function, and the reconstruction of the
magnetization curve is straightforward. The slopes of the two pieces
correspond to the two critical fields $\hc{1}$ and $\hc{2}$. Below $\hc{1}$, the
magnetization is identically zero ($\Mag = 0$), it jumps at $\hc{1}$ to half
the saturation value ($\Mag = 1/2$), remains constant up to $\hc{2}$ and then
jumps again to the saturation value ($\Mag = 1$). This behaviour has already
been predicted in Ref. \cite{Mila}
on the basis of a strong coupling analysis. The corresponding
transition fields $\hc{1}$ and $\hc{2}$ are computed easily
from Eq.\ (\ref{Eexact}). One finds that
\beq
\hc{1} = \Jp \, , \qquad\qquad
\hc{2} = 2 J + \Jp \, .
\eeq{hc1-2}

If on the contrary $\Jp$ is small enough, $E_{tt}$ is always below $E_{st}$, and
the magnetization curve is identical to that of a spin-1 chain. In particular,  
it raises smoothly between $\hc{1}$ and $\hc{2}$ and has no discontinuity at
these points. 

In the intermediate region, the situation is far more complicated because the
two curves intersect several times. Increasing $\Jp$ from small values, $E_{st}$
first touches $E_{tt}$ at $\Mag = 1/2$ for $\Jp/J=1.3807(5)$. Beyond but close
enough to that value, there will thus be two intersections below and above 
$\Mag = 1/2$. However this
is not the end of the story since the energy given by $\min(E_{tt},E_{st})$ \
is no longer convex. So one has to perform Maxwell constructions on each side
of the point $\Mag = 1/2$. They are shown as dashed
lines in Fig.\ \ref{energy2}. The slopes of the energy on each side of 
$\Mag = 1/2$ then give
the two critical fields between which a plateau $\Mag = 1/2$ exists. At both
critical fields there will be a discontinuity in the magnetization since the
slope of the energy is constant over a finite range of magnetization.

\begin{figure}[t]
\psfig{figure=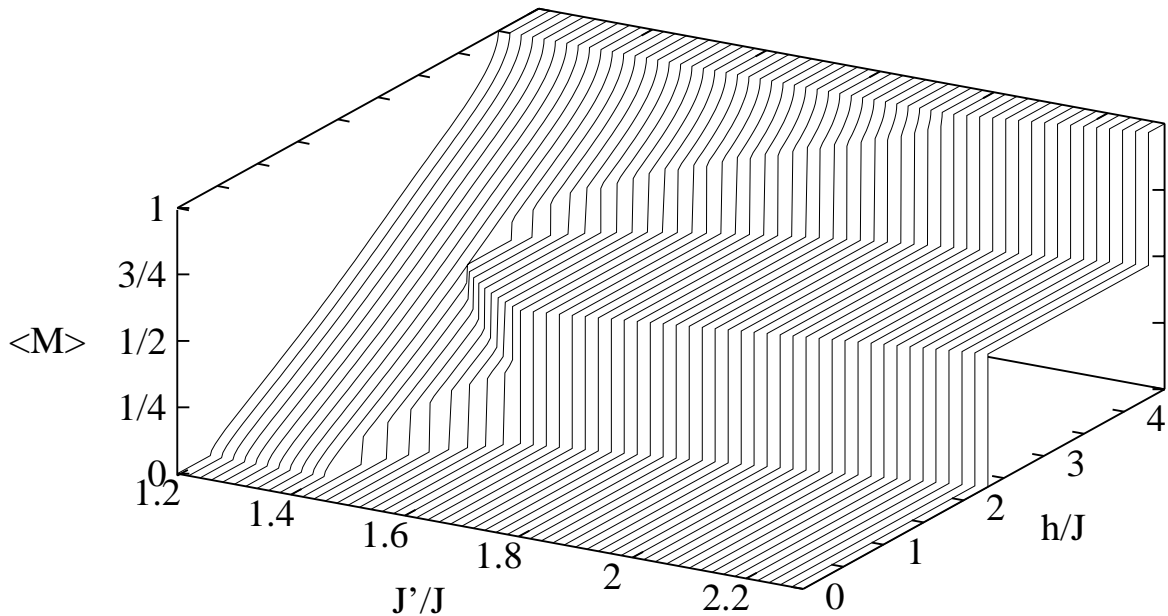,width=\columnwidth,angle=270}
\smallskip
\caption{
Magnetization curves of the two-leg ladder Eq.\ (\ref{Hop}).
\label{mcurve2}
}
\end{figure}

Increasing $\Jp$ further, another transition occurs where $E_{st}$ touches
$E_{tt}$ at $\Mag=0$, \ie\ when $\Jp/J=-e_\infty=1.401484...$
\footnote{The critical values of $\Jp$ for $\Mag = 0$ and $\Mag = 1/2$
are surprising close but can be distinguished safely with our
accuracy.
}.
Beyond that value, a Maxwell construction is again necessary resulting in a
discontinuity of the magnetization at the corresponding critical field.

\begin{figure}[t]
\psfig{figure=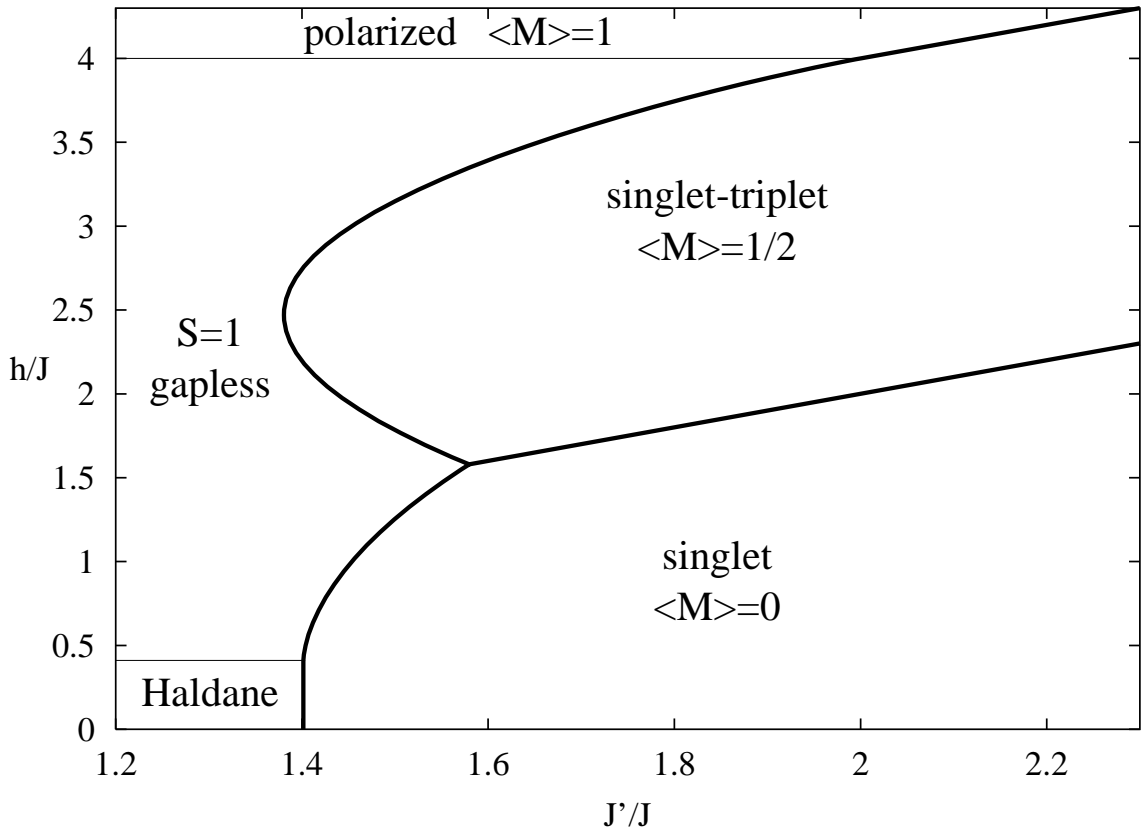,width=\columnwidth,angle=270}
\smallskip
\caption{
Ground state phase diagram for the two-leg ladder. The heavy lines
denote first order transitions, the thin ones second order transitions.
\label{p2-diag}
}
\end{figure}

The above conclusions are illustrated by Fig.\ \ref{mcurve2} which shows
the evolution of the magnetization curves with $\Jp$
obtained from the DMRG data.
It is also straightforward to construct
the full ground state phase diagram of the two-leg ladder as a function
of $\Jp/J$ and $h$ -- see Fig.\ \ref{p2-diag}.

For $h/J < \Delta_{\infty}$ and $\Jp/J < -e_{\infty}$, the ground state
is the Haldane gap ground state \cite{Haldane} with a gap to magnetic
excitations.  
For larger magnetic fields one finds that the ground state is given
by the corresponding one of the $S=1$ chain. The magnetization curve
in this region is smoothly varying demonstrating the presence of
gapless magnetic excitations. In this phase, the $\Mag = 1/2$
plateau opens at $\Jp/J = 1.3807(5)$, $h/J = 2.4706(2)$.
Typically, the opening of a plateau as a function of
a coupling constant would have to occur via a Kosterlitz-Thouless
transition. Here it is clearly of a different type due to the
fact that the opening of the $\Mag = 1/2$ plateau occurs 
by a crossing of the energy levels Eq.\ (\ref{Eexact}) and Eq.\ (\ref{Ett})
at $S^z=L/2$.

The transitions between the $S=1$ gapless phase and the $\Mag = 0$
and $1/2$ plateaux are first order transitions as a function of the
external magnetic field. In the ($\Jp/J$,$\Mag$)--plane one would
therefore find finite regions in the phase diagram where the system
phase-separates into regions with finite $S=1$ chains and regions
which singlets on all rungs (for $\Mag < 1/2$) or alternating
singlets and polarized triplets (for $\Mag > 1/2$).

The $S=1$ gapless
phase, the singlet $\Mag = 0$ phase and the $\Mag = 1/2$ plateau
meet at $h/J = \Jp/J = 1.5796(4)$.

\section{Three legs}

For the Hamiltonian of Eq.\ (\ref{Hop}) with $N=3$ the relevant eigenstates are:
\begin{enumerate}
\item
Spin-3/2 states on all rungs with energy
\beq
E_{3/2} = {3 \over 4} \Jp L + J E^{S=3/2}(L,S^z) \, ,
\eeq{E3o2}
\item
alternating spin-1/2 and -3/2 on the rungs with corresponding
energy
\beq
E_{3/2-1/2} = J E^{S=3/2-1/2}(L,S^z) \, ,
\eeq{E3o2-1o2}
\item
spin-1/2 on each rung with energy
\beq
E_{1/2} = -{3 \over 4} \Jp L + J E^{S=1/2}(L,S^z) \, .
\eeq{E1o2}
\end{enumerate}
Here $E^{\bullet}(L,S^z)$ is the energy of the corresponding spin-chain
with coupling constant $1$, length $L$ and a given $S^z$-sector.
It should be noted that each spin-1/2 state comes with two chiralities.
The eigenvalues of the Hamiltonian of Eq.\ (\ref{Hop}) are independent
of these chiralities, which is clear from the rewriting of Eq.\ (\ref{Heff}). 
This degeneracy gives rise to an entropy $\ln(2)$ for each rung with spin 1/2.

\begin{figure}[t]
\centerline{\psfig{figure=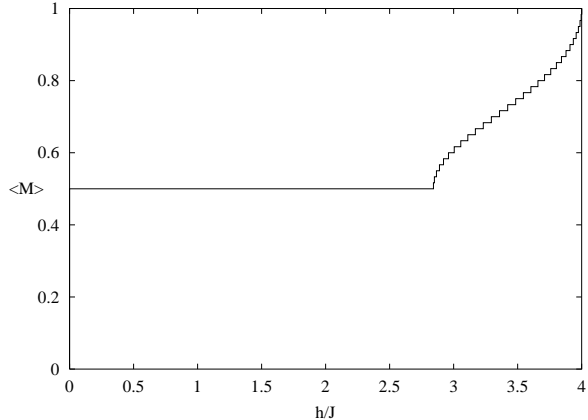,width=0.5\columnwidth,angle=270}}
\smallskip
\caption{
Magnetization curve of the $S=3/2-1/2$ ferrimagnetic chain for $L=60$.
\label{m-ferri}
}
\end{figure}

On ladders with $L \le 8$ (a total of up to 24 sites) we have again checked
numerically that the ground states in the presence of a magnetic field
can always be found among
Eq.\ (\ref{E3o2})--(\ref{E1o2}) except for special values of the field where the
magnetization jumps. Then we have again used DMRG to compute
$E^{S=3/2}(60,S^z)$ and $E^{S=3/2-1/2}(60,S^z)$. 
Since we are not aware of any previous discussion of the magnetization curve
of the $S=3/2-1/2$ ferrimagnetic chain, we present it in Fig.\ \ref{m-ferri}.
This curve is very similar to the magnetization curve of the
$S=1-1/2$ ferrimagnetic chain \cite{Kuramoto,MSBMY}, the main
difference being the plateau value of $\Mag$ which is $1/3$ in the latter
case.

$E^{S=1/2}(\infty,S^z=L \Mag /2)/L$ was obtained from the Bethe
ansatz equations for the thermodynamic limit in the spirit
of \cite{Griffiths} (the actual program used is a small modification
of the one described in \cite{CHPPRB}). The accuracy of our DMRG
results can be assessed by comparison to earlier DMRG studies.
We find $E^{S=3/2}(60,0)/60 \approx -2.82879$ and
$E^{S=3/2-1/2}(60,0)/60 \approx -0.98362$ which should
be compared to the extrapolated values
$e_{\infty}^{S=3/2} = -2.82833(1)$ \cite{HWHM} and
$e_{\infty}^{S=3/2-1/2} = -0.98362$ \cite{PRS}, respectively
\footnote{Our result for the antiferromagnetic gap
$\Delta_{L} = E^{S=3/2-1/2}(L,L/2+1) - E^{S=3/2-1/2}(L,L/2)$ is 
$\Delta_{60} \approx 2.8420$. This cannot be directly compared
to the corresponding result of \cite{PRS} since that DMRG
computation was performed for open boundary conditions and
in this case a {\it bound state} with the boundary is formed.
Repetition of our computation with open boundary conditions
lead to $\Delta_{60}^{(o)} \approx 1.8558$ which compares well
with the extrapolated value $\Delta_{\infty}^{(o)} = 1.8558(1)$ \cite{PRS}.
Also for open boundary conditions,
the next magnetic excitation lies above the antiferromagnetic gap, \ie\
it can indeed be expected that the magnetization process becomes independent
of the boundary conditions in the thermodynamic limit.
}.

\begin{figure}[t]
\centerline{\psfig{figure=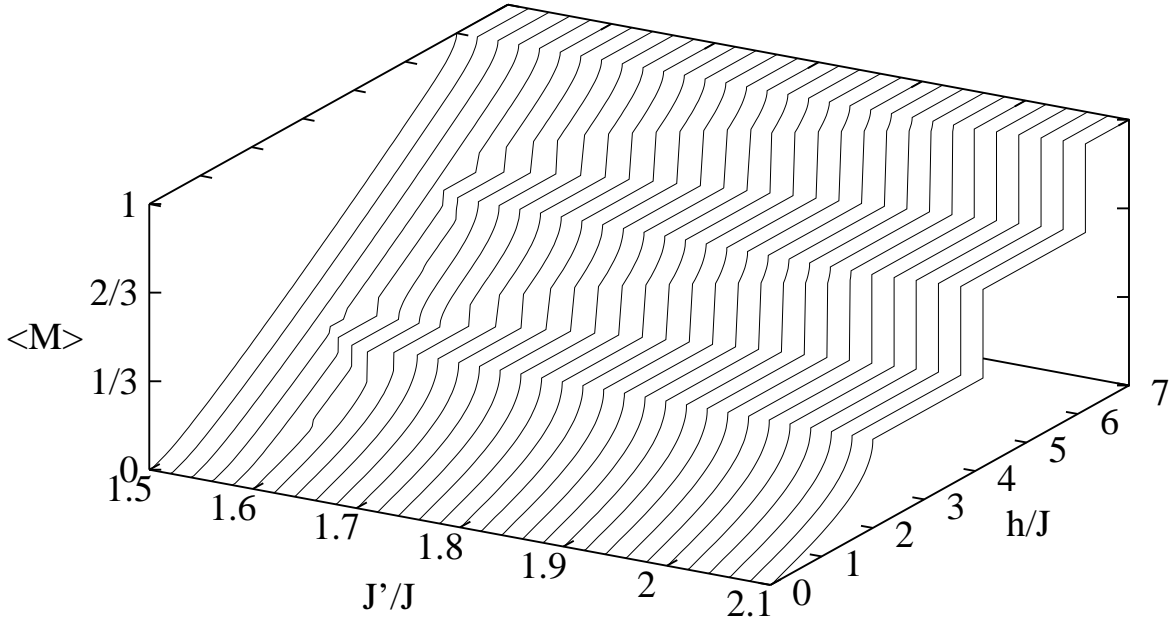,width=\columnwidth,angle=270}}
\smallskip
\caption{
Magnetization curves of the three-leg ladder (Eq.\ (\ref{Hop})).
\label{mcurve3}
}
\end{figure}

\begin{figure}[t]
\psfig{figure=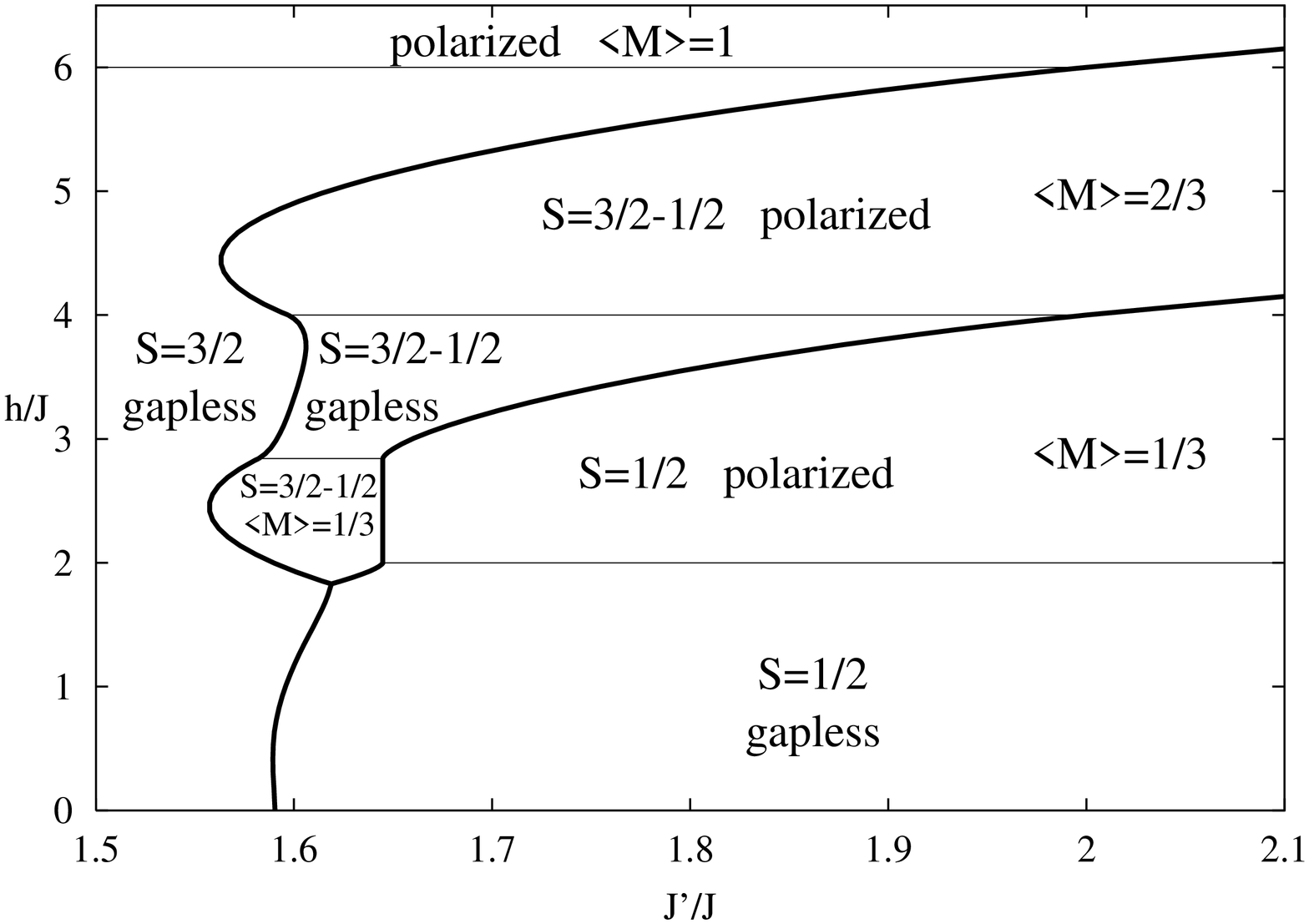,width=\columnwidth,angle=270}
\smallskip
\caption{
Ground state phase diagram for the three-leg ladder. The heavy lines
denote first order transitions, the thin ones second order transitions.
\label{p3-diag}
}
\end{figure}

Proceeding as for the two-leg ladder, the magnetization curves
of Eq.\ (\ref{Hop}) for $N=3$ 
can be constructed from these data for different values of $\Jp/J$. 
The evolution with $\Jp/J$ is
shown in Fig.\ \ref{mcurve3}. Figure\ \ref{p3-diag} shows a projection
onto the $\Jp/J$-$h$ plane. Construction of this ground state phase diagram is
somewhat more involved than Fig.\ \ref{p2-diag} and slightly less accurate.
We therefore do not quote any number for points in Fig.\ \ref{p3-diag},
but all transitions should be accurate to the order of the
width of the lines in the figure.

For $\Jp \ale 1.557 J$, the magnetization curve of the three-leg ladder
is identical to that of the $S=3/2$ chain. If $\Jp \ge 2 J$,
the magnetization process proceeds by polarizing $S=1/2$ states
at $\hc{1} = 2 J$. Then the magnetization jumps from the polarized
$S=1/2$ state ($\Mag = 1/3$ in the language of the three-leg ladder)
to the polarized state of the ferrimagnetic $S=3/2-1/2$ chain
($\Mag = 2/3$ in the language of the three-leg ladder) at
\beq
\hc{2} = J + {3 \over 2} \Jp \, .
\eeq{hc2p3}
Finally, the magnetization jumps again polarizing the complete system
at
\beq
\hc{3} = 3 J + {3 \over 2} \Jp
\eeq{hc3p3}
in this region $\Jp \ge 2 J$.

For intermediate $\Jp$ partially polarized states of the
ferrimagnetic $S=3/2-1/2$ chain also contribute to the magnetization
process, and the plateau state which has $\Mag = 1/2$
in the language of the $S=3/2-1/2$ chain appears. When translated into the
language of the three-leg ladder the latter leads to $\Mag = 1/3$ and it is
this number which we quote in the corresponding region of Fig.\ \ref{p3-diag}.
At $\Jp \approx 1.645 J$, there is a first order transition between this
plateau state and the fully polarized state of the $S=1/2$ chain. Since
the magnetization is the same in both states, there is no jump in the
magnetization. Still, the transition occurs via a level crossing and
it is therefore first order in the sense that many correlators on the
$\Mag = 1/3$ plateau are discontinuous across this line.

A remarkable property of the Hamiltonian of Eq.\ (\ref{Hop}) with $N=3$ is 
that it
has a plateau at $\Mag = 2/3$ without giving rise to a gap (or plateau)
at $\Mag = 0$. On general grounds both of them would be permitted
for a frustrated $N=3$-leg ladder when translational invariance is
spontaneously broken in the ground state to a period $p=2$
\cite{CHPPRL,CHPPRB,tandon,CHPzigzag}. The present situation should be
contrasted to the case of the regular cylindrical three-leg ladder
\cite{CHPPRL,CHPPRB} which has a plateau at $\Mag = 0$ (\ie\ a gap
\cite{Schulz,Arrigoni,KaTa}) and presumably no plateau at $\Mag = 2/3$
\cite{COAIQ}. In that case, the
plateau at $\Mag = 0$ is a direct consequence of the coupling of the spin and
the chirality of each triangular rung. In the present case, there is no plateau
because the chirality is completely decoupled from the spin.
We would also like to emphasize that there is a residual entropy both
on the $\Mag = 1/3$ and on the $\Mag = 2/3$ plateaux of $\ln(2)$ and $\ln(2)/2$,
respectively.

\section{Conclusion}

We have proposed and studied a class of frustrated ladders for which the 
magnetization curve can be calculated by elementary methods once the
magnetization of a few, much simpler systems is known. There are many further
models in this class beyond the one studied in the present paper. For
example, the models studied in \cite{KTS} and \cite{MTM} come to mind
as natural generalizations of the two- and three-leg ladders which we
have studied. A modified version of the model of \cite{KTS} has in fact
been proposed to describe the $S=1/2$ trimer system
Cu$_3$Cl$_6$(H$_2$O)$_2 \cdot$2H$_8$C$_4$SO$_2$ \cite{trimer,OTTK}.
Since the synthesis and investigation of many quasi-one dimensional magnets
is under way, also the models discussed here should soon become relevant
to experimental systems.

A remark is in order regarding a small detuning of the coupling constants
from the case where the reasoning of the present paper applies. After
such a detuning, the spins on each rung are no longer conserved
and therefore one will find avoided crossings
rather than real crossings. Since the plateau-state is gapped,
magnetization plateaux are stable against small perturbations but
a softening of the transitions between them is to be expected.
This can indeed be seen easily in the strong-coupling region
$\Jp \gg J,\Jd$ where magnetization plateaux can be shown to
exist using perturbative arguments and transitions between them
can be described by first-order effective Hamiltonians (see e.g.\
\cite{Totsuka,CHPPRL,CHPPRB,Mila,CHABetal,tandon,COAIQ,WeHa}).
With the choice $J = \Jd$ one eliminates the kinetic energy part
of the first-order effective Hamiltonian and therefore we find
steps in the magnetization curves even at finite $J=\Jd < 2 \Jp$ whereas
for $J \ne \Jd$ one would find a smooth transition. However,
the fact that we found jumps in these special
models is still interesting since it points towards the possibility 
of steep increases of the magnetization in frustrated models in general, a
possibility not emphasized so far in that context.

The main advantage of the models Eq.\ (\ref{Hop}) is that the features
of the magnetization curve - plateaux and
jumps - can be traced back to level crossings. The underlying physical picture 
is thus clear and simple, and at the same time a very precise determination of 
the critical fields is possible. The price to pay is not horrendous - the
Hamiltonian can still be written down compactly, and its visualization is
straightforward. So, in the spirit of the Affleck-Kennedy-Lieb-Tasaki model of a
spin-1 chain with a gap \cite{AKLT}, we hope that these models will be a
useful reference to understand the physics of the magnetization of
low-dimensional magnets.

\subsection*{Note added}
After completion of this work, we became aware of \cite{MSKU} which
uses similar ideas to compute the magnetization curve of a two-dimensional
model.

\acknowledgments
We are indebted to S.R.\ White for help with the DMRG calculations.
F.M.\ is grateful to the Institut f\"ur Theoretische Physik of the
ETH Z\"urich for hospitality during the course of this project.

\end{document}